\documentclass[journal,10pt,twosided]{IEEEtran}

\usepackage{amsmath}
\usepackage{amssymb}
\usepackage{graphicx}
\usepackage{pgfplots}
\usepackage{algorithm}
\usepackage{algpseudocode}
\usepackage{cite}
\usepackage{subfigure}
\usepackage{balance} % balancing a page by calling \balance
\usetikzlibrary{shapes,snakes}
\usetikzlibrary{calc}
\usetikzlibrary{patterns}
\usetikzlibrary{decorations.pathmorphing} 

\usepackage[normalem]{ulem}
\usepackage{soul}

\newtheorem{remark}{Remark}

\newcommand\redsout{\bgroup\markoverwith{\textcolor{red}{\rule[0.5ex]{2pt}{0.4pt}}}\ULon}

\newcommand{\Ali}{
\textcolor{black}
}
\newcommand{\RxPowerSplitting}[4]{
\coordinate (a) at (#1,#2);
\draw[line width=0.25pt,scale=(#3)] (a)--($(a)+(-0.2,0)$)--($(a)+(-0.2,0.7)$)--
($(a)+(-0.1,0.8)$)--($(a)+(-0.3,0.8)$)--($(a)+(-0.2,0.7)$);
\draw ($(a)+(0.15,0)$) circle (0.15cm);
\node at ($(a)+(0.15,0)$) {\small +};
\draw ($(a)+(0.3,0)$)--($(a)+(0.5,0)$);
\draw[->] ($(a)+(0.15,-0.5)$)--($(a)+(0.15,-0.15)$);
\node at ($(a)+(0.15,-0.7)$) {$w_{#4}$};
\draw[fill=yellow!20] ($(a)+(0,-0.2)+(0.5,0)$) rectangle ($(a)+(1,0.2)+(0.5,0)$);
\node at ($(a)+(0.5,0)+(0.5,0)$){\small $\text{PS}_{#4}$};
}
\newcommand{\TxAntenna}[3]{
\coordinate (a) at (#1,#2);
\draw[line width=0.25pt,scale=(#3)] (a)--($(a)+(0.2,0)$)--($(a)+(0.2,0.7)$)--
($(a)+(0.1,0.8)$)--($(a)+(0.3,0.8)$)--($(a)+(0.2,0.7)$);
\draw[fill=green!20!red!20] ($(a)+(0.2,0)+(-1.4,-0.2)$) rectangle($(a)+(0.2,0)+(-0.2,0.2)$); 
\node at ($(a)+(0.2,0)+(-0.8,0)$){TX};
}

\newcommand{\Aydin}{
\textcolor{black}
}

\begin{document}
\title{Optimal\Ali{Power Splitting} for Simultaneous Information Detection and Energy Harvesting}
\author{
\IEEEauthorblockN{Ali Kariminezhad, Soheil Gherekhloo, and Aydin Sezgin}\\
\normalsize{Institute of Digital Communication Systems}\\
\normalsize{RUB, 44801 Bochum, Germany}\\
\small{Email: \{ali.kariminezhad, soheyl.gherekhloo, aydin.sezgin\}@rub.de}
}
\maketitle

\begin{abstract}
\Aydin{This letter deals with the joint information and energy processing at a receiver of a point-to-point communication channel. In particular, the trade-off between the achievable information rate and harvested energy for a multiple-antenna\Ali{power splitting} (PS) receiver is investigated.
Here, the rate-energy region characterization is of particular interest, which is intrinsically a non-convex problem. In this letter, an efficient algorithm is proposed for obtaining an approximate solution to the problem in polynomial time. This algorithm is mainly based on the Taylor approximation in conjunction with semidefinite relaxation (SDR) which is solved by interior-point methods. Moreover, we utilize the Gaussian randomization procedure to obtain a feasible solution for the original problem. It is shown that by proper receiver design the rate-energy region can be significantly enlarged compared to the state of the art, while at the same time the receiver hardware costs is reduced by utilizing less number of energy harvesting circuitry.} 
\end{abstract}
\section{Introduction}
\begin{figure}
\centering
\begin{tikzpicture}[scale=0.6, every node/.style={scale=0.8}]
\
\TxAntenna{-4}{-1.75}{1}
\RxPowerSplitting{-2}{0}{1}{1}
\node at (-2.25,-1.6){.};
\node at (-2.25,-1.75){.};
\node at (-2.25,-1.9){.};

\node at (-1,-1.6){.};
\node at (-1,-1.75){.};
\node at (-1,-1.9){.};
\RxPowerSplitting{-2}{-3.5}{1}{K}

\draw (-0.5,0.1)--(0,0.8);
\draw (0,0.8)--(2,0.8);
\node at (1,1.1){$\sqrt{\lambda_1}$};
\draw (2.2,0.8) circle (0.2);
\node at (2.2,0.8){$+$};
\draw[->] (2.2,1.5)--(2.2,1);
\node at (2.2,1.7){$n_1$};
\draw (2.4,0.8)--(3.4,0.8);
\draw (3.6,0.8) circle (0.2);
\node at (3.6,0.8){$\times$};
\draw[->] (3.6,1.5)--(3.6,1);
\node at (3.6,1.7){$u^{*}_1$};
\draw (3.8,0.8)--(4.8,0.8);
\draw [->](4.8,0.8)--(5.7,-0.8);
\node at (5.4,-0.85){.};
\node at (5.4,-0.95){.};
\node at (5.4,-1.05){.};
\draw (6,-0.95) circle (0.3);
\node at (6,-0.95){$\large +$};
\draw[->] (6.3,-0.95)--(7.5,-0.95);
\node at (6.9,-0.75){$y_{\text{ID}}$};
\draw[fill=red!20] (7.5,-1.25) rectangle (9.1,-0.65);
\node at (8.3,-0.95){ID};

\draw (-0.5,-0.1)--(0,-0.8);
\draw (0,-0.8)--(2.7,-0.8);
\node at (1.25,-1.1){$\sqrt{1-\lambda_1}$};
\draw (2.9,-0.8) circle (0.2);
\node at (2.9,-0.8){$\times$};
\draw[->] (2.9,-1.5)--(2.9,-1);
\node at (2.9,-1.7){$v^{*}_1$};
\draw (3.1,-0.8)--(4.8,-0.8);
\draw[->] (4.8,-0.8)--(5.7,-2.4);
\node at (5.4,-2.45){.};
\node at (5.4,-2.55){.};
\node at (5.4,-2.65){.};
\draw (6,-2.55) circle (0.3);
\node at (6,-2.55){$\large +$};
\draw[->] (6.3,-2.55)--(7.5,-2.55);
\node at (6.9,-2.2){$y_{\text{EH}}$};
\draw[fill=green!20] (7.5,-2.85) rectangle (9.1,-2.25);
\node at (8.3,-2.55){EH};

\draw (-0.5,-3.4)--(0,-2.7);
\draw (0,-2.7)--(2,-2.7);
\node at (1,-2.4){$\sqrt{\lambda_K}$};
\draw (2.2,-2.7) circle (0.2);
\node at (2.2,-2.7){$+$};
\draw[->] (2.2,-2)--(2.2,-2.5);
\node at (2.2,-1.8){$n_K$};
\draw (2.4,-2.7)--(3.4,-2.7);
\draw (3.6,-2.7) circle (0.2);
\node at (3.6,-2.7){$\times$};
\draw[->] (3.6,-2)--(3.6,-2.5);
\node at (3.6,-1.8){$u^{*}_K$};
\draw (3.8,-2.7)--(4.8,-2.7);
\draw[->] (4.8,-2.7)--(5.7,-1.1);

\draw (-0.5,-3.6)--(0,-4.3);
\draw (0,-4.3)--(2.7,-4.3);
\node at (1.25,-4.6){$\sqrt{1-\lambda_K}$};
\draw (2.9,-4.3) circle (0.2);
\node at (2.9,-4.3){$\times$};
\draw[->] (2.9,-5)--(2.9,-4.5);
\node at (2.9,-5.3){$v^{*}_K$};
\draw (3.1,-4.3)--(4.8,-4.3);
\draw[->] (4.8,-4.3)--(5.7,-2.7);

% channels
\draw[dashed,->] (-3.5,-1.6)--(-2.4,0);
\node[rotate=60] at (-3,-0.4){$h_1$};
\draw[dashed,->] (-3.5,-1.9)--(-2.4,-3.4);
\node[rotate=-50] at (-2.7,-2.6){$h_K$};
\end{tikzpicture}
\caption{Joint information detection (ID) and energy harvesting (EH) in SIMO-P2P by power splitting (PS). The receiver is equipped with $K$ antennas. Power splitting coefficient at $k$th antenna is shown by $\lambda_k$.}
\label{fig:SystemModel}
\end{figure}
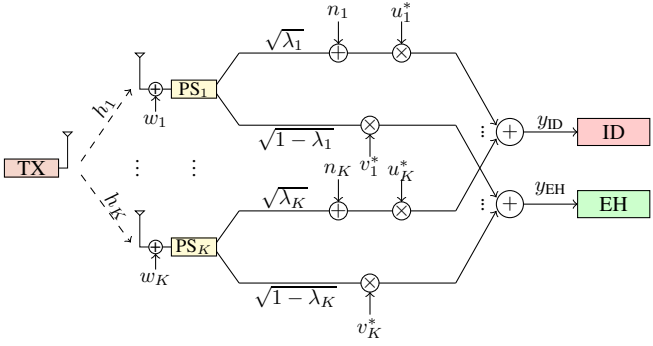

\Aydin{Recent trends in communications indicate the involvement of the energy demands of the customers besides their information demands~\cite{Krikidis2014},~\cite{Liu2015}.} Thus, power transfer along with the classical information transfer has to be considered \Ali{jointly for optimal transceiver design.}\\
\Ali{Though energy can be harvested from different energy sources, this letter considers radio frequency (RF) signals as the energy source~\cite{Lu2015}. Thereby, a rectifier is required for converting the incident RF energy to\Ali{DC power}, which in turn charges the energy buffer. This allows the users to stay operational in the network for a longer time. The authors in~\cite{Hameed2014,Bolos2016,Jabbar2010} study the RF-DC conversion chain where the power conversion efficiency is addressed.}\\
Recently, simultaneous wireless information and power transfer (SWIPT) has been studied from different perspectives in various communication scenarios.
The authors in~\cite{Amor2016} characterize the fundamental limits of simultaneous information and power transmission in a two-user Gaussian multiple access channel, however~\cite{Zhang2013},~\cite{Huang2013} investigate SWIPT in a downlink channel.\Ali{Besides, the analysis of SWIPT in sensor networks was recently addressed in~\cite{Pan2017}.} Moreover,~\cite{Kariminezhad2016} studies{\Ali a} two-tier network from the achievable rate-energy perspective, where the benefits of improper Gaussian signaling for joint information detection (ID) and energy harvesting (EH)\Ali{are} highlighted. Having simultaneous information and power transmission, joint ID and EH structures need to be exploited at the receivers. Time sharing (TS)~\cite{Ni2016} and\Ali{power splitting} (PS) are two potential schemes for this purpose in single antenna receivers, while antenna switching (AS) is\Ali{another} structure that can be applied in multi-antenna receivers~\cite{Krikidis2014}.\Aydin{With PS, the received signal power at each antenna is split into two portions, one undergoes the ID processing chain and the other undergoes the EH circuitry. In comparison, the TS structure utilizes the whole received signal power for the information purpose in one time slot and for energy purpose in the other time slot. With AS, the information is captured from a subset of antennas, while the energy is\Ali{harvested by} the\Ali{complementary} subset.}{\Ali Power splitting} is the optimal receiver structure in multiple-antenna receivers, since it includes AS as a special case, Fig.~\ref{fig:SystemModel}.\Ali{The authors in~\cite{Cai2016} investigate the joint transceiver design in MISO relay systems for optimized ID and EH purposes.} The authors in~\cite{Shi2014} consider power splitting structure in MISO systems, where the optimal beamforming design is investigated.\Ali{Moreover,~\cite{Zhang2017} studies MIMO power splitting design under a mean-square error (MSE) constraint}. Although the authors in~\cite{Liu2013} study the optimal power splitting in a single-input multiple-output point-to-point (SIMO-P2P) channel, they neglect beam-steering for constructively adding up the energy signals for the EH purpose. Furthermore,~\cite{Liu2013} ignores the antenna noise for simplicity.\\

\textbf{Our Contribution:} In this letter we consider:
\begin{itemize}
\item the exact capacity and received energy of the SIMO-P2P channel including antenna noise,
\item \Ali{separate} beamforming for both EH and ID purposes,
\end{itemize}
which render the problem to be non-convex. We propose an efficient algorithm to solve the problem in polynomial time.\Ali{Eventually,} the performance of the proposed algorithm is compared with exhaustive search solution for evaluation purposes.
%{\Ali
%\subsection{Notation} Throughout the paper, we represent vectors in boldface lower-case letters while the matrices are expressed in boldface upper-case. ${\rm{Tr}}(\bf{A})$, ${\bf{A}}^{T}$, ${\bf{A}}^{H}$ and ${\bf{A}}^{*}$ represent the trace, transpose, Hermitian and complex conjugate of matrix $\mathbf A$, respectively. Moreover,  $\text{eig}\left({\bf{A}}\right)$ is the eigenvalue operator and $\bf{A}\circ \bf{B}$ represents the Hadamard product between $\bf{A}$ and $\bf{B}$. The operator $\text{diag}(.)$ generates a diagonal matrix if the argument is a vector. Moreover, it generates a vector of diagonal elements if the argument is a matrix.}

%\textcolor{red}{\textbf{Notations:} $a^*$, $A^H$, $\mathbf{I}_r$, $\text{diag}$, $a_k$ represents the $k$th element of vector $\mathbf{a}$, 
%$\boldsymbol{0}$, $\boldsymbol{1}$.}

\section{System Model}
We consider a single-input multiple-output (SIMO) point-to-point (P2P) channel with $K$ antennas at the receiver side. The channel input-output relationship can be modelled as
\begin{align}
\mathbf{r}= \mathbf{h}s+\mathbf{w},
\end{align} 
where $\mathbf{r}\in \mathbb{C}^{K\times 1}$ represents the received signal vector and its $k$th element, i.e., $r_k$ is the received signal at the $k$th antenna. 
Moreover, the variable $s$ is the transmit signal with $\mathbb{E}\{|s|^2\} = P$, $\mathbf{w}\in \mathbb{C}^{K\times 1}$ denotes the received circularly symmetric additive white Gaussian noise (AWGN) vector at the receiver antenna array \Aydin{(e.g., antenna mismatch noise)} with $\mathbb{E}\{\mathbf{w}\mathbf{w}^H\} =\mathbf{\Sigma}_w^2=\sigma_w^{2}\mathbf{I}_K$, and $\mathbf{h}\in\mathbb{C}^{K\times 1}$ is the channel vector,\Ali{ which is assumed to be known at the receiver}.

In this system, the receiver is not only interested in the information of the received signal but also its energy.
Hence, as it is shown in Fig.~\ref{fig:SystemModel}, the receiver splits the received signal at each antenna into two signals by using a power splitter (PS).
These signals are further used for information detection (ID) and energy harvesting (EH) purposes. For instance, at the $k$th antenna, the received signal $r_k$ is split into $r_{\text{ID},k} = \sqrt{\lambda_k} r_k$ and $r_{\text{EH},k} =\sqrt{1-\lambda_k} r_k$, where $\lambda_k\in[0,1]$ is the power splitting coefficient used for the $k$th received signal which needs to be further optimized. 
While $r_{\text{ID},k}$ is used for the ID chain, the receiver harvests the energy of $r_{\text{EH},k}$.
Here, we suppose that the receiver processes a noisy version of $r_{\text{ID},k}$ given by 
$\tilde{r}_{\text{ID},k} = r_{\text{ID},k} + n_k,$ 
where $n_k$ denotes the processing noise \Aydin{(e.g., mixer noise, amplification noise, etc.)} of the $k$th ID chain which is circularly symmetric AWGN with zero-mean and variance $\sigma_{n}^2$. 
Furthermore, it is assumed that $n_k$ is independent of all other variables. 
In what follows, we present the information detection and energy harvesting process separately.

\textbf{Information detection:} Here, the receiver combines first its observations \Ali{$\tilde{r}_{\text{ID},1},\ldots,\tilde{r}_{\text{ID},K}$} linearly to obtain 
\begin{align}
y_{\text{ID}} = \sum_{k=1} ^K u_k^* \tilde{r}_{\text{ID},k}&{\Ali =\sum_{k=1} ^K u_k^* \sqrt{\lambda_k} h_k s + \sum_{k=1}^K u_k^* \left(\sqrt{\lambda_k} w_k + n_k \right)} \notag \\ 
&= \mathbf{u}^H  \mathbf{L}^{1/2} \mathbf{h} s + \mathbf{u}^H (\mathbf{L}^{1/2}\mathbf{w} + \mathbf{n})
 \label{eq:ID_received_sig}
\end{align}
where $u_k^*$ is the complex-valued weighting coefficient for $\tilde{r}_{\text{ID},k}$, $\mathbf{u} = \begin{bmatrix}
u_1 &...&u_K
\end{bmatrix}^{T}
$, $\mathbf{L}=\text{diag}\left(\lambda_1,\ldots,\lambda_K \right)$, and $\mathbf{n}=\begin{bmatrix}
n_1 &...&n_K \end{bmatrix}^{T}$.\Ali{Notice that $(.)^{*}$ is the complex conjugate operator}. In fact, the receiver decodes its desired symbol $s$ from $y_{\text{ID}}$ with the following communication rate,
\begin{align}
R(\mathbf{u},\boldsymbol{\lambda})= \log\left(1+ \frac{\mathbf{u}^H \mathbf{L}^{1/2} \mathbf{h} \mathbf{h}^H \mathbf{L}^{1/2} \mathbf{u} P}{\mathbf{u}^H \left(\mathbf{L} \mathbf{\Sigma}_w^2 + \mathbf{\Sigma}_n^2\right)  \mathbf{u}} \right),\label{eq:Rate1}
\end{align}
where $\boldsymbol{\lambda}=\begin{bmatrix}
\lambda_1 &...&\lambda_K
\end{bmatrix}^{T}$ and $\mathbf{\Sigma}_n = \sigma_n^{2}\mathbf{I}_K$.
The maximum reliable information rate can be achieved by 
\begin{align}
\mathbf{u}_{\text{opt}}=\text{eig}\left((\mathbf{L} \mathbf{\Sigma}_w^2 + \mathbf{\Sigma}_n^2)^{-1} \mathbf{L}^{1/2} \mathbf{h} \mathbf{h}^H \mathbf{L}^{1/2}\right),\label{BFvector}
\end{align}
\Ali{where eig(.) is the eigenvector operator. By beamforming in the direction of $\mathbf{u}_{\text{opt}}$, the following information rate is achievable~\cite{Liu2013},}
\begin{align}
\bar{R}(\boldsymbol{\lambda}) &= \log\left(1+ \mathbf{h}^H \mathbf{L}^{1/2} \left(\mathbf{L} \mathbf{\Sigma}_w^2 + \mathbf{\Sigma}_n^2\right)^{-1} \mathbf{L}^{1/2} \mathbf{h} P\right)\notag \\
&= \log\left(1+ \sum_{k=1}^K\frac{\lambda_k |h_k|^2P}{\lambda_k \sigma_{w_k}^2 + \sigma_{n_k}^2} \right). \label{eq:Rate}
\end{align}

\textbf{Energy harvesting:}
First, the receiver generates a linear superposition of the complex-valued signals $r_{\text{EH},k}$ with $k=1\ldots,K$, as\Ali {$
y_{\text{EH}} = \sum_{k=1}^K v_k^* r_{\text{EH},k},
$}
where $v_k$ is a complex-valued weighting coefficient with unit magnitude.
The receiver harvests the energy of $y_{\text{EH}}$\Ali{which is defined by $\tau \mathbb{E}\{y_{\text{EH}}y^{*}_{\text{EH}}\}$ where $\tau$ denotes the energy conversion loss in converting the harvested RF energy into DC power. This can be formulated as}
\begin{align}
\hspace*{-0.25cm} E(\mathbf{v},\boldsymbol{\lambda}) =\tau P \left|\sum_{k
=1}^K\sqrt{1-\lambda_k} v_k^* h_k\right|^2   + \tau\sum_{k
=1}^K(1-\lambda_k)\sigma_{w_k}^2, \label{Harv_Energy}
\end{align}
where  $\mathbf{v} = \begin{bmatrix}
v_1 &...&v_K \end{bmatrix}^{T}$.
By choosing $v_k = \frac{h_k}{|h_k|}$, the expression in \eqref{Harv_Energy} is maximized due to Cauchy-Schwarz inequality. 
Hence, the maximum harvested energy is given by 
\begin{align}
\hspace*{-0.25cm} \bar{E}(\boldsymbol{\lambda}) = \tau P \left(\sum_{k
=1}^K\sqrt{1-\lambda_k} |h_k|\right)^2   + \tau \sum_{k
=1}^K(1-\lambda_k)\sigma_{w_k}^2, \label{eq:energy}
\end{align}
which is the achievable energy considering single energy conversion chain\Ali{after constructive beamforming}.
\section{Rate-energy Region}
\Ali{One can easily conclude from \eqref{eq:Rate} and \eqref{eq:energy} that the achievable rate and harvested energy depends on the value of $\boldsymbol{\lambda}$. 
In the extreme cases when $\boldsymbol{\lambda} = \boldsymbol{0}$, the receiver harvests maximal energy of the received signal $\mathbf{r}$ while the achievable rate is zero. 
On the other hand, by setting $\boldsymbol{\lambda} = \boldsymbol{1}$, the receiver uses the received signal only for information detection, hence we obtain the maximum achievable rate. 
In other words, there is a trade-off between the achievable rate and harvested energy. 
This trade-off can be described by the so-called rate-energy region. 
This region is the set of all rate-energy pairs which are achievable using the presented receive strategy.
Formally, the rate-energy pair is defined as follows
\begin{align*}
\small{\mathcal{S}=\{\left(R(\boldsymbol{\lambda}),E(\boldsymbol{\lambda})\right)|R(\boldsymbol{\lambda}) \leq \bar{R}(\boldsymbol{\lambda}),E(\boldsymbol{\lambda}) \leq \bar{E}(\boldsymbol{\lambda}), \mathbf{0}\leq\boldsymbol{\lambda}\leq \boldsymbol{1}\}.}
\end{align*}
The main goal of this work is to characterize the rate-energy region $\mathcal{S}$. To this end, in what follows, we formulate an optimization problem which gives us the outer-most boundary of the rate-energy region.
\begin{remark}
The authors in~\cite{Liu2013} consider multiple energy conversion chains (one chain per receive antenna). Thus, the total harvested energy is
\begin{align}
\hat{E}(\boldsymbol{\lambda}) = \tau P \sum_{k
=1}^K (1-\lambda_k) |h_k|^{2}  + \tau\sum_{k
=1}^K(1-\lambda_k)\sigma_{w_k}^2. \label{eq:energy2}
\end{align}
This structure is more costly compared to the structure utilized in this paper. Moreover, the total harvested energy in~\eqref{eq:energy2} is a lower-bound for~\eqref{eq:energy}.
\end{remark}}
In this section we formulate an appropriate optimization problem for characterizing the outer-most boundary of the rate-energy region.
we formulate the energy maximization problem under rate constraint as
\begin{subequations}\label{A1}
\begin{align}
\max_{\lambda_k,\ \forall k}\quad& \bar{E}(\boldsymbol{\lambda})\tag{\ref{A1}}\\
\text{subject to}\quad &\bar{R}(\boldsymbol{\lambda})\geq\psi,\quad
\Ali{\mathbf{0}\leq \boldsymbol{\lambda}\leq \mathbf{1}},
\end{align}
\end{subequations}
where $\psi$ is the rate demand. By increasing the rate constraint up to $\bar{R}(\mathbf{1})$ (i.e., $\psi\in [0,\bar{R}(\mathbf{1})]$) and solving the problem, the rate-energy region can be obtained. The objective function in \eqref{A1} is a non-convex function due to the summation of square roots in~\eqref{eq:energy}. Moreover, the rate constraint with the achievable rate stated in~\eqref{eq:Rate} yields a non-convex set. Hence, problem \eqref{A1} is a non-convex problem. Here, we ignore antenna noise (i.e., $\mathbf{w}$) in the beamforming phase in \eqref{BFvector}. Notice that, antenna noise is only neglected in the beamforming phase, but not on the rate expression in~\eqref{eq:Rate1}. Therefore, the following beamforming vector is utilized for ID purpose,
\begin{align}
\Ali{\mathbf{\tilde{u}}_{K\times 1}}=\mathbf{L}^{1/2}\mathbf{h}.\label{BFvector2}
\end{align}
In order to formulate the achievable rate we define the given vectors
\begin{align}
\Ali{\hat{\boldsymbol{\lambda}}_{K\times 1}}=&\left[
\sqrt{1-\lambda_1}\ \ ...\ \ \sqrt{1-\lambda_K}
\right]^{T},\label{a2}\\
\mathbf{g}_1=&\left[
|h_1|^{2} \ \ ...\ \ |h_K|^{2} 
\right]^{T},\
\mathbf{g}_2=\left[
|h_1| \ \ ... \ \ |h_K| 
\right]^{T},\label{a3}\\
w^{'}_k=& h^{*}_kw_k ,\quad\quad\quad
n^{'}_k= h^{*}_kn_k .\label{a4}
\end{align}
Note that, the equivalent variances of the antenna and processing noise are $\sigma^{2}_{w^{'}_k}=|h_k|^{2}\sigma^{2}_{w}$ and $\sigma^{2}_{n^{'}_k}=|h_k|^{2}\sigma^{2}_{n}$, respectively. Furthermore, we define the matrices
\begin{align}
\mathbf{G}_1=\mathbf{g}_1\mathbf{g}_1^{T},\quad \mathbf{G}_2=P\mathbf{g}_2\mathbf{g}_2^{T}+{\sigma}^{2}_w\mathbf{I},\label{G12}
\end{align}
Now, by using the definitions in~\eqref{a2}-\eqref{G12}, we reformulate the achievable information rate in \eqref{eq:Rate1} utilizing \eqref{BFvector2} and the harvested energy in \eqref{eq:energy} as
\begin{align}
\tilde{R}=&\log_2\left(1+\frac{P\boldsymbol{\lambda}^{T}
\mathbf{G}_1\boldsymbol{\lambda}}{\boldsymbol{\lambda}^{T}\boldsymbol{\Sigma}\boldsymbol{\lambda}+
\boldsymbol{\lambda}^{T}\boldsymbol{\sigma}}\right),\label{RateNew}\\
\bar{E}=&\tau \hat{\boldsymbol{\lambda}}^{T}\mathbf{G}_2\hat{\boldsymbol{\lambda}},\label{EnergyNew}
\end{align}
respectively. Here, the vector of equivalent processing noise variances and equivalent antenna noise covariance matrix are represented by
\begin{align}
\boldsymbol{\sigma}=&\begin{bmatrix}
\sigma^{2}_{n^{'}_1}& ... & \sigma^{2}_{n^{'}_K}
\end{bmatrix}^{T},\quad \boldsymbol{\Sigma}=&\text{diag}(
\sigma^{2}_{w^{'}_1},...,\sigma^{2}_{w^{'}_K}),
\end{align}
respectively. Now, we approximate $\hat{\boldsymbol{\lambda}}$ by using the Taylor approximation. Then, each element of $\hat{\boldsymbol{\lambda}}$, i.e., $\sqrt{1-\lambda_k}$, around $a_k$ can be approximated upto the first polynomial order as
\begin{align}
\sqrt{1-\lambda_k}\approx (1-a_k)^{0.5}-\frac{1}{2}(1-a_k)^{-0.5}(\lambda_k-a_k).\label{taylorApprox}
\end{align}
It is important to note that, this approximation is not defined at $\lambda_k=1,\ \forall k$, hence, we include this case by simply calculating the achievable rate for $\boldsymbol{\lambda}=\mathbf{1}$ to characterize the rate-energy region completely. Furthermore, it is necessary to restrict the solution region within a small gap around approximation point, (i.e., $a_k-\epsilon\leq\lambda_k\leq a_k+\epsilon,\ \forall k$). This restriction guarantees the Taylor approximation accuracy.\Aydin{By plugging~\eqref{taylorApprox} into~\eqref{a2}, we obtain}
\begin{align}
\hat{\boldsymbol{\lambda}}\approx\boldsymbol{\alpha}+\frac{1}{2}\left(\boldsymbol{\beta}-\tilde{\boldsymbol{\lambda}}\right),\label{LambdaHat}
\end{align}
where 
\begin{align}
\boldsymbol{\alpha}=&\begin{bmatrix}
(1-a_1)^{0.5}& ...& (1-a_K)^{0.5}
\end{bmatrix}^{T},\\
\boldsymbol{\beta}=&\begin{bmatrix}
a_1(1-a_1)^{-0.5}& ...& a_K(1-a_K)^{-0.5}
\end{bmatrix}^{T},\\
\tilde{\boldsymbol{\lambda}}=&\begin{bmatrix}
\lambda_1(1-a_1)^{-0.5}& ...& \lambda_K(1-a_K)^{-0.5}
\end{bmatrix}^{T}.
\end{align}
\Aydin{Now, substituting ~\eqref{LambdaHat} into~\eqref{EnergyNew}, the harvested energy can be approximated by}
\begin{align}
\bar{E}\approx \tau \left(\Gamma+\frac{1}{4}\tilde{\boldsymbol{\lambda}}^{T}\mathbf{G}_2
\tilde{\boldsymbol{\lambda}}-\left(\boldsymbol{\alpha}+\frac{1}{2}\boldsymbol{\beta}\right)^{T}\mathbf{G}_2
\tilde{\boldsymbol{\lambda}}\right),
\end{align}
where the offset $\Gamma$ (independent of the optimization parameter) is\Ali{ 
$
\Gamma=\boldsymbol{\alpha}^{T}\mathbf{G}_2\boldsymbol{\alpha}+
\frac{1}{4}\boldsymbol{\beta}^{T}\mathbf{G}_2\boldsymbol{\beta}+
\boldsymbol{\alpha}^{T}\mathbf{G}_2\boldsymbol{\beta}.
$}
The relation between $\tilde{\boldsymbol{\lambda}}$ and $\boldsymbol{\lambda}$ can be easily seen to be 
$\tilde{\boldsymbol{\lambda}}=\mathbf{M}\boldsymbol{\lambda}$
where
$\mathbf{M}=\text{diag}\left(
\begin{bmatrix}
(1-a_1)^{-0.5} & ... & (1-a_K)^{-0.5}
\end{bmatrix}^{T} \right)$.
Thus, 
\begin{align}
\bar{E}(\boldsymbol{\lambda})\approx \tau\left(\Gamma+\frac{1}{4}\boldsymbol{\lambda}^{T}\mathbf{M}^{T}\mathbf{G}_2
\mathbf{M}\boldsymbol{\lambda}-\left(\boldsymbol{\alpha}+\frac{1}{2}\boldsymbol{\beta}\right)^{T}\mathbf{G}_2
\mathbf{M}\boldsymbol{\lambda}\right).
\end{align}
\begin{figure*}[t]
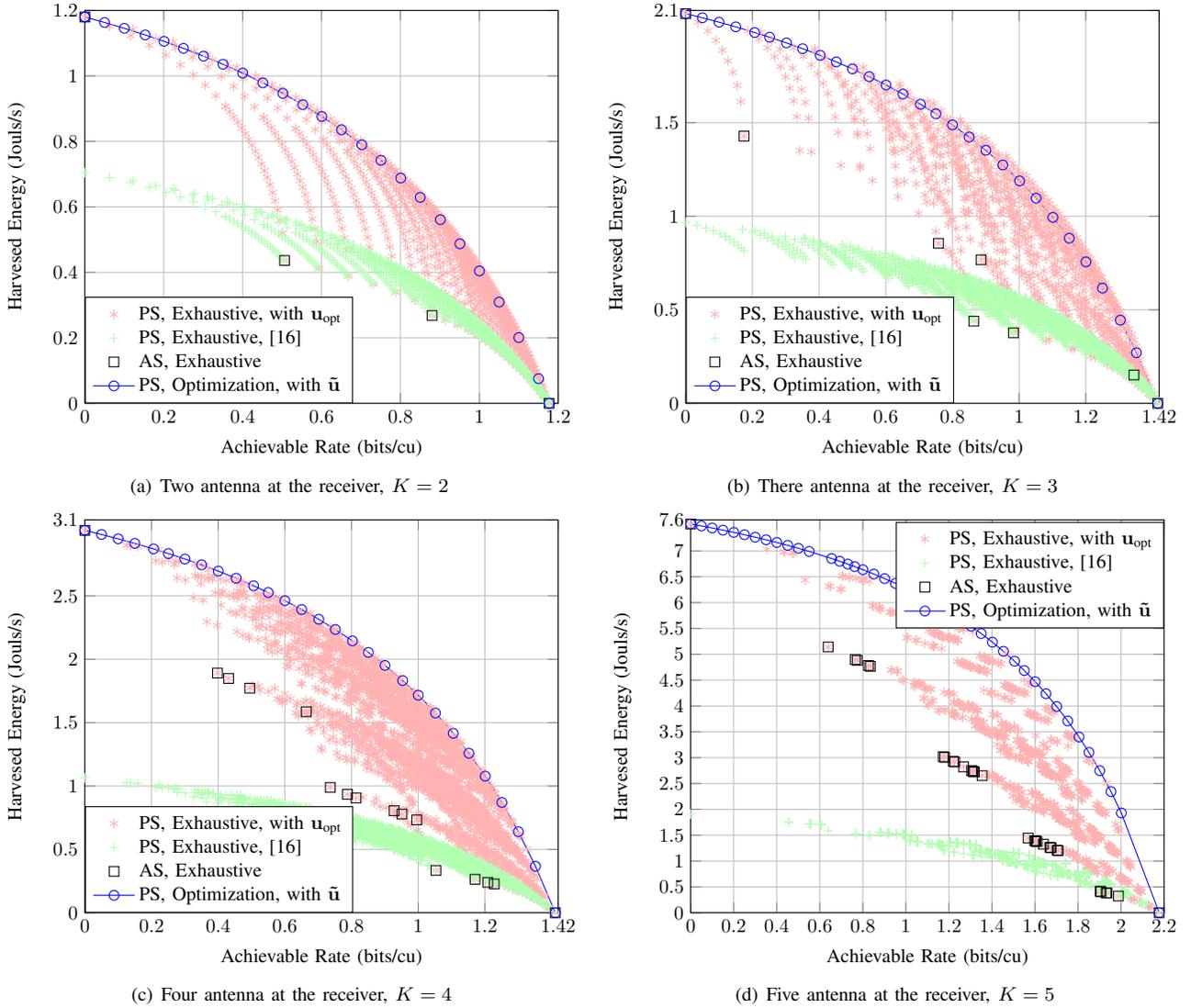

\centering
\subfigure[Two antenna at the receiver, $K=2$]{
\tikzset{every picture/.style={scale=1}, every node/.style={scale=0.8}}%
\input{A1}
\label{fig:a1}
}
\subfigure[There antenna at the receiver, $K=3$]{
\tikzset{every picture/.style={scale=1}, every node/.style={scale=0.8}}%
\input{C1}
\label{fig:c1}
}
\subfigure[Four antenna at the receiver, $K=4$]{
\tikzset{every picture/.style={scale=1}, every node/.style={scale=.8}}%
\input{B1}
\label{fig:b1}
}
\subfigure[Five antenna at the receiver, $K=5$]{
\tikzset{every picture/.style={scale=1}, every node/.style={scale=0.8}}%
\input{D1}
\label{fig:d1}
}
\caption{Comparison between\Ali{power splitting} (PS) and antenna switching (AS) in single-input multiple-output point-to-point (SIMO-P2P) channel. The proposed algorithm is able to almost capture the outer-most rate-energy boundary.}
\label{fig:MMU}
\end{figure*}
For convenience, we define $\mathbf{G}^{'}_2=\mathbf{M}^{T}\mathbf{G}_2
\mathbf{M}$, $\mathbf{G}^{''}_2=\mathbf{G}_2
\mathbf{M}$ and $\boldsymbol{\zeta}=\left(\boldsymbol{\alpha}+\frac{1}{2}\boldsymbol{\beta}\right)$. Now, having the approximate of the harvested energy, we formulate problem \eqref{A1} as a semidefinite program (SDP) as
\begin{subequations}\label{A2}
\begin{align}
\max_{\boldsymbol{\lambda},\boldsymbol{\Lambda}}\quad& \tau \left(\Gamma+\frac{1}{4}\text{Tr}(\boldsymbol{\Lambda}\mathbf{G}^{'}_2)
-\boldsymbol{\zeta}^{T}\mathbf{G}^{''}_2
\boldsymbol{\lambda}\right)\tag{\ref{A2}}\\
\text{s.t.}\quad & P\text{Tr}(\boldsymbol{\Lambda}
\mathbf{G}_1) - (2^{\psi}-1)\left( \text{Tr}( {\boldsymbol{\Lambda}\boldsymbol{\Sigma})+
\boldsymbol{\lambda}^{T}\boldsymbol{\sigma}}\right)\geq 0,\\
& \mathbf{0}\leq \boldsymbol{\lambda}\leq \mathbf{1},\label{region11}\\
& \boldsymbol{\Lambda}=\boldsymbol{\lambda}\boldsymbol{\lambda}^{T}.\label{rankCons1}
\end{align}
\end{subequations}
Problem \eqref{A2} is a non-convex SDP due to the rank constraint in \eqref{rankCons1},~\cite{Boyd2004}. We relax this problem by allowing $\boldsymbol{\Lambda}\succeq\boldsymbol{\lambda}\boldsymbol{\lambda}^{T}$ and dropping the rank-1 constraint. By the Schur complement, we know
\begin{align}
\boldsymbol{\Lambda}-\boldsymbol{\lambda}\boldsymbol{\lambda}^{T}
\succeq 0
\rightarrow
\begin{bmatrix}
\boldsymbol{\Lambda} & \boldsymbol{\lambda} \\
\boldsymbol{\lambda}^{T} & 1
\end{bmatrix}\succeq 0.\label{replace1}
\end{align}
By replacing constraint \eqref{rankCons1} by \eqref{replace1}, the resulting semidefinite relaxation (SDR) can be efficiently solved~\cite{Luo2010}.\Ali{To tighten the relaxation~\eqref{replace1}, we use the fact that~\eqref{taylorApprox} is accurate for $\lambda_k$ around $a_k$ and the optimal $\lambda_k$ is obtained by reducing the gap between $\lambda_k$ and $a_k$ iteratively, i.e., $a^{\star}_k=\lambda^{\star}_k$. Thus, we consider the additional side constraint $\Lambda_{kk}=a_k \lambda_k,\ \forall k$, where $\Lambda_{kk}$ is the $k$th diagonal element of the auxiliary variable $\boldsymbol{\Lambda}$. This side constraint can be reformulated as $\text{diag}\left( \boldsymbol{\Lambda} \right)=\mathbf{a}^{(l)}\circ \boldsymbol{\lambda}$, where $\circ$ represents the Hadamard product. The solution of the resulting SDR (i.e., $\boldsymbol{\lambda}^{\star},\boldsymbol{\Lambda}^{\star}$) is not guaranteed to be a feasible solution for the original problem, since either $\text{rank}\left(\boldsymbol{\Lambda}^{\star}\right)\neq 1$, or $\boldsymbol{\Lambda}^{\star}\neq \boldsymbol{\lambda}^{\star}\boldsymbol{\lambda}^{\star^{T}}$}. Here, we use Gaussian randomization to acquire a good sub-optimal solution by generating random vectors with Gaussian  distribution as $\mathcal{CN}\sim \left(\boldsymbol{\lambda}^{\star},\boldsymbol{\Lambda}^{\star}\right)$~\cite{Huang2014}. Then we solve the problem as described in algorithm 1.

\begin{algorithm}
\caption{Optimal power splitting}
\begin{algorithmic}[1]
\State \Ali{calculate $\bar{R}(\mathbf{1})$ and fix information rate $0 \leq \psi \leq \bar{R}(\mathbf{1})$,}
\State \Ali{initialize $\mathbf{a}^{(0)}=\frac{\psi}{\bar{R}(\mathbf{1})}$,}
\State $l=0$ (iteration index),
\While{$E^{(l)}-E^{(l-1)}$ large}
\State determine $\boldsymbol{\epsilon}^{(l)}=\boldsymbol{\eta}\min(\mathbf{a}^{(l)})$, where $\boldsymbol{\eta} < \mathbf{1}$,
\State  Solve the following SDR,
\begin{subequations}\label{A3}
\begin{align}
\max_{\boldsymbol{\lambda},\boldsymbol{\Lambda}}\quad& \Gamma+\frac{1}{4}\text{Tr}(\boldsymbol{\Lambda}\mathbf{G}^{'}_2)
-\boldsymbol{\zeta}^{T}\mathbf{G}^{''}_2
\boldsymbol{\lambda}\tag{\ref{A3}}\\
\text{s.t.}\quad & P\text{Tr}(\boldsymbol{\Lambda}
\mathbf{G}_1) - (2^{\psi}-1)\left( \text{Tr}( {\boldsymbol{\Lambda}\boldsymbol{\Sigma})+
\boldsymbol{\lambda}^{T}\boldsymbol{\sigma}}\right)\geq 0,\label{rateCons22}\\
& \mathbf{a}^{(l)}-\boldsymbol{\epsilon}^{(l)}\leq \boldsymbol{\lambda}\leq \mathbf{a}^{(l)}+\boldsymbol{\epsilon}^{(l)},\label{region22}\\
& \begin{bmatrix}
\boldsymbol{\Lambda} & \boldsymbol{\lambda} \\
\boldsymbol{\lambda}^{T} & 1
\end{bmatrix}\succeq 0,\label{rankCons2}\\
&\Ali{\text{diag}\left( \boldsymbol{\Lambda} \right)=\mathbf{a}^{(l)}\circ \boldsymbol{\lambda} }.
\end{align}
\end{subequations}
\State Generate $N$ Gaussian randomizations with distribution $\boldsymbol{\lambda}_i\sim\mathcal{CN}(\boldsymbol{\lambda}^{*},\boldsymbol{\Lambda}^{*}),\ \forall i\in\{1,...,N\}$,
\State Force $\boldsymbol{\lambda}_i$ to be in $[\mathbf{a}^{(l)}-\boldsymbol{\epsilon}^{(l)},\mathbf{a}^{(l)}+\boldsymbol{\epsilon}^{(l)}]$,
\State $j=1$,
\If {$\boldsymbol{\lambda}_i$ fulfils constraint \eqref{rateCons22}, where $\boldsymbol{\Lambda}_i=\boldsymbol{\lambda}_i
\boldsymbol{\lambda}^{T}_i$}
\State $\boldsymbol{\lambda}^{'}_j=\boldsymbol{\lambda}_i$ and  $\boldsymbol{\Lambda}^{'}_j=\boldsymbol{\lambda}_i\boldsymbol{\lambda}_i^{T}$, potential solutions, 

\State Calculate $E^{'}_j$ from~\eqref{eq:energy}.
\State $j=j+1$.
\EndIf
\State Optimal solution: $\boldsymbol{\lambda}^{''}=\text{arg}\max \mathcal{E}^{'}$,\\
\hspace*{0.5cm}$E^{(l)}=\tau \max \mathcal{E}^{'}$ where $\mathcal{E}^{'}=\{E^{'}_j,\ \forall j\}$,
\State $l=l+1$,
\State $\mathbf{a}^{(l)}=\boldsymbol{\lambda}^{''}$.
\EndWhile
\end{algorithmic}
\end{algorithm}
\section{Numerical Results}
In this section, we provide the numerical solutions which is based on algorithm 1. Here, we consider two and four receive antennas with either antenna switching (AS) or\Ali{power splitting} (PS) receivers for simultaneous ID and EH purposes. Transmit signal power is $2$ watts, while antenna and processing noise are $\mathcal{CN}(0,0.1)$. Furthermore, the energy conversion loss is assumed to be $\tau = 1$. For the sake of reproducibility of the results, we provide the simulated channels for the two upto five antenna cases as,
\begin{align}
\mathbf{h_1}=&[0.41e^{i0.95},  \  0.29e^{i1.44}]^{T},\\
\mathbf{h_2}=&[0.37e^{i0.42},\  0.42e^{i1.4}, \ 0.16e^{i0.78} ]^{T},\\
\mathbf{h_3}=&[0.26e^{i0.12},  \  0.29e^{i2.15},  \  0.34e^{i1.80},   \ 0.24e^{i2.08}]^{T},\\
\mathbf{h_4}=&[0.37e^{i0.79},  \  0.33e^{i0.43},  \  0.38e^{i0.12},\nonumber\\
& \ 0.40e^{i0.85} ,   \ 0.39e^{i1.05}]^{T},
\end{align} 
As depicted in Fig.~\ref{fig:MMU}, the proposed algorithm delivers the outer-most rate-energy tuples, meanwhile performing almost as good as exhaustive search. As a benchmark, the efficiency of PS structure is compared with AS and the performance gap is highlighted. Moreover, as shown in Fig.~\ref{fig:MMU}, the studied PS structure in combination with beamforming promises enlarged rate-energy region compared to the structure investigated in~\cite{Liu2013}.
\Ali{\section{Conclusion}
In this letter, we proposed an efficient energy harvesting scheme for the multiple-antenna receiver applying\Ali{power splitting} structure. In this scheme, the received signal undergoes an energy harvesting chain after constructive\Ali{power splitting} and beamforming. The\Ali{power splitting} problem turns out to be a non-convex optimization problem. As a remedy, a polynomial-time algorithm to obtain an efficient solution is proposed. The proposed power splitting structure along with the algorithm can be utilized in SWIPT for various wireless communication networks and near-field car-to-car communications. Moreover, the energy harvesting sensors in wireless sensor networks could benefit from this scheme in order to stay operational for a longer time. }

\bibliographystyle{IEEEtran} 
\bibliography{reference}
\balance

\end{document}